\documentclass[aps,pra,twocolumn,10pt]{revtex4-1}

\usepackage{graphicx,amsmath,amssymb}
\usepackage{color}
\def\EE#1{\times 10^{#1}}

\begin{document}
\hbadness = 10000

\title{A hexapole-compensated magneto-optical trap on a mesoscopic atom chip}

\author{S. J\"ollenbeck$^1$, J. Mahnke$^1$, R. Randoll$^1$, W. Ertmer$^1$, J. Arlt$^2$, C. Klempt$^1$}
\affiliation{$^1$ Institut f\"ur Quantenoptik, Leibniz Universit\"at Hannover, 30167~Hannover, Germany}
\affiliation{$^2$ QUANTOP, Danish National Research Foundation Center for Quantum Optics, Department of Physics and Astronomy, Aarhus University, 8000 \AA{}rhus C, Denmark}

\date{\today}

\begin{abstract}
Magneto-optical traps on atom chips are usually restricted to small atomic samples due to a limited capture volume caused primarily by distorted field configurations. Here we present a magneto-optical trap with minimized distortions based on a mesoscopic wire structure which provides a loading rate of $8.4\EE{10}$~atoms/s and a maximum number of $8.7\EE{9}$ captured atoms. The wire structure is placed outside of the vacuum to enable a further adaptation to new scientific objectives. Since all magnetic fields are applied locally without the need for external bias fields, the presented setup will facilitate parallel generation of Bose-Einstein condensates on a conveyor belt with a cycle rate above $1$~Hz.

\end{abstract}

\maketitle

\section{Introduction}

The field of atom interferometry with laser-cooled atoms has quickly evolved within the recent past. In principle, the sensitivity of these interferometers can be improved further by using quantum degenerate input states~\cite{Wang2005,Bouyer1997}. However it is not advantageous to use a standard apparatus for the production of atomic Bose-Einstein condensates (BECs) as a source for an atom interferometer. The advantages of the coherent input state are outweighed by a severe loss in both atom number and cycle rate. Whereas typical atom interferometers operate with $10^8$ atoms and a cycle time of $100$~ms~\cite{Treutlein2001}, a conventional BEC apparatus with a standard magnetic trap can produce samples of $10^5 -10^7$ atoms within $10-60$ seconds. Two constraints are limiting the total flux of these atom sources: Firstly, the magneto-optical trap~(MOT) as the first cooling step has a limited loading rate due to the finite capture range and a maximum trappable atom number due to density dependent light-induced losses~\cite{Sesko1989}. Secondly, the creation of BECs relies on evaporative cooling, which is limited in speed by the available atomic densities that are necessary for the continuous thermalization process~\cite{Ketterle1996}.

A number of techniques are available to achieve the high atomic flux from a MOT that is needed for the creation of BECs in quick succession. The MOT loading rate can be increased by the use of cold atomic beams provided by a Chirp~\cite{Ertmer1985} or Zeeman slower~\cite{Phillips1982}, or an efficient two-dimensional MOT~\cite{Dieckmann1998,Schoser2002}. Furthermore, the MOT size can be increased by using high laser power, large trapping beams and particularly uniform quadrupole fields. These methods have been employed to capture up to $10^{11}$ Sodium atoms in $1$~s~\cite{Streed2006} and $10^{10}$ Rubidium atoms in $100$~ms~\cite{Chaudhuri2006}. To achieve high cycle rates it is also important to reduce the evaporative cooling time to obtain BEC. This  can be achieved in tightly confining traps, best realized on microscopic chip traps~\cite{Reichel2002,Folman2002,Fortagh2007} or in optical dipole traps~\cite{Barrett2004}. Both methods feature fast evaporation, but typically yield small atom numbers. Optical traps suffer particularly from poor mode matching with the MOT. Atom chips provide better mode matching since they can provide both the magnetic trapping field and the MOT's quadrupole field. However, the MOTs on these chips yield small atom numbers due to the small trapping volume of the strongly distorted quadrupole field. It has been shown that the distortions can be decreased~\cite{Wildermuth2004}, however large atom numbers have not been achieved in a pure chip trap. 

In this paper we present a mesoscopic chip MOT that captures $8.7\EE{9}$~atoms with an initial loading rate of  $8.4\EE{10}$~atoms/s without the need for any external fields. The MOT's quadrupole field is generated by millimeter-scale wires~\cite{Wildermuth2004} that are mounted on a planar surface. This is a complementary approach to most atom chip arrangements~\cite{Reichel2002,Folman2002,Fortagh2007}. The wire arrangement creates an extended quadrupole field with compensated hexapole component. This allows for a large volume of the MOT, and thus ensures large captured atom numbers. To keep maximal flexibility for further modifications of the chip it is mounted outside the vacuum system. It is separated from the atoms by a gold coated steel foil which forms part of the vacuum system. Moreover, all required magnetic fields are generated on the chip itself, avoiding the need for an external bias field often provided by large additional coils. 

The exceptionally high loading rate achieved in this system in combination with its mesoscopic chip structure makes it an ideal source for the production of BECs at a high cycle rate. In the future, this system will allow for serial production of BECs and thus presents a source capable of delivering BECs with a cycle rate above $1$~Hz. In this sense, the apparatus presents a complementary approach to the continuous evaporation in a cold atomic beam~\cite{Lahaye2004}.

This paper is organized as follows. In section~\ref{atomchip}, we describe the mesoscopic atom chip with special emphasis on its design and the hexapole compensation. Section \ref{setup} outlines the full experimental setup. The results obtained with the mesoscopic chip MOT are presented in section \ref{results} and section~\ref{outlook} concludes with a summary and an outlook towards the high-flux BEC source.

\section{Mesoscopic atom chip}
\label{atomchip}

The goal of creating a high-flux source of laser cooled atoms leads to a number of requirements for the design of the chip. First of all, the quadrupole field should only be generated by local wires on the chip, without the need for external bias fields. Any external field generated by large coils would impede the independent adjustment of the local magnetic fields for the transport of the atomic clouds. Furthermore the chip surface has to be at a reasonable distance from the center of the MOT to allow for a large trapping volume. The trapping volume can be increased further by reducing the distortions of the magnetic quadrupole field, i.e. by compensating the field's hexapole component. Finally, the need for large currents in the mesoscopic wires makes it desirable to place the atom chip outside the vacuum system. Thus a large number of vacuum feedthroughs can be avoided and efficient cooling of the wires can easily be realized. Moreover this arrangement allows for rapid changes to the chip structure without compromising the vacuum inside the experimental chamber.

\subsection{Hexapole compensated quadrupole field}

The simplest design to realize a two-dimensional quadrupole field is a single current carrying wire supplemented with an external bias field~\cite{Folman2002}. However, this arrangement suffers from severe distortions of the quadrupole field even at moderate distances from its center. 

\begin{figure}[ht]
	\centering
		\includegraphics[width=\columnwidth]{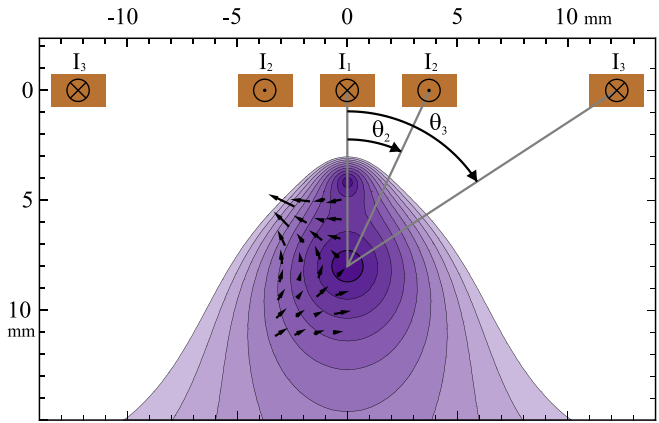}
		\caption{Spatial structure of the hexapole compensated magnetic quadrupole field. The copper rectangles represent a cut through the wire structure and the angles $\theta_2$ and $\theta_3$ are indicated.}
	\label{fig:2d-quad}
\end{figure}

To obtain a large trapping volume it is necessary to compensate these distortions. The most dominant contribution to these distortions is the hexapole component of the field. By adding two additional parallel wires, it is possible to compensate the hexapole component~\cite{Hyodo2007}. Moreover the bias field can be generated by another two parallel coplanar wires, replacing the need for external bias coils. Figure~\ref{fig:2d-quad} shows the necessary wire arrangement. At the center of the trap, all of these wires have dipole, quadrupole and hexapole contributions which have to be included in the analysis. Hence our design aims to minimize both the dipole (shifts the center of the trap) and the hexapole components of the field while maximizing the quadrupole component.

For this optimization, we follow the notation of Ref.~\cite{Hyodo2007}, where the angles $\theta_2$ and $\theta_3$ indicate the angles of the lines connecting the position of the wires with the center of the trap as shown in Fig.~\ref{fig:2d-quad}. The multipole components $\eta_\text{l}$ can be given in the form

\begin{eqnarray}
\eta_\text{l} &=& I_1 + \nonumber \\
&\phantom{=}& 2\; I_2\; \text{cos}(\theta_2)^{l+1} \; \text{cos}\left[(l+1)\theta_2 \right] + \nonumber \\
&\phantom{=}& 2\; I_3\; \text{cos}(\theta_3)^{l+1} \; \text{cos}\left[(l+1)\theta_3 \right]
\end{eqnarray}

where $I_1$, $I_2$, and $I_3$ represent the currents through the center wire and the inner and the outer wire pair respectively. The center current $I_1$ can be treated as a general scaling parameter which determines the mean gradient of the quadrupole field. For a given angle $\theta_2$, the angle $\theta_3$ can be optimized to yield a maximum quadrupole component under the condition that $I_2$ and $I_3$ are chosen such that the dipole and the hexapole component are zero. 

\begin{figure}[ht]
	\centering
		\includegraphics[width=\columnwidth]{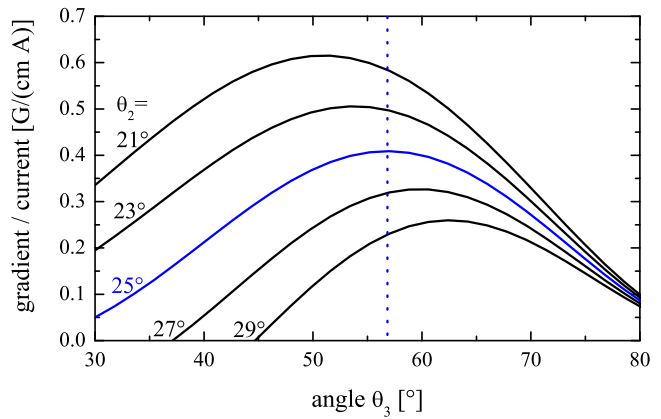}
		\caption{Optimization of the wire structure. For each angle $\theta_2$, the gradient for a given current is shown as a function of $\theta_3$. This allows for the optimal choice of $\theta_2$ and $\theta_3$ under the experimental conditions.}
	\label{fig:2d-opt}
\end{figure}

Figure~\ref{fig:2d-opt} shows the result of such an optimization for angles $\theta_2$ from $21^\circ$ to $29^\circ$. For the experiment, an angle of $\theta_2=25^\circ$ was chosen to be able to place the wires at a comfortable distance from each other and work with an optimized angle $\theta_3=57^\circ$. The ratio between the currents is given by $I_2=-1.21\ I_1$ and $I_3=1.65\ I_1$ to obtain an optimized quadrupole field. The resulting field configuration is shown in Fig~\ref{fig:2d-quad}. The indicated lines of equal magnetic field strength form circles up to a diameter of $4$~mm. In addition the direction of the magnetic field vectors is crucial for the operation of a MOT. In the chosen configuration the magnetic field vectors are parallel to the incident laser beams for a minimum diameter of $8$~mm, providing a large trapping volume of the MOT. Note that the hexapole compensation creates a second field minimum above the center of the MOT. Due to its inverse orientation, it presents an antibinding quadrupole field which limits the size of the trapping region.

\begin{figure}[ht]
	\centering
		\includegraphics[width=\columnwidth]{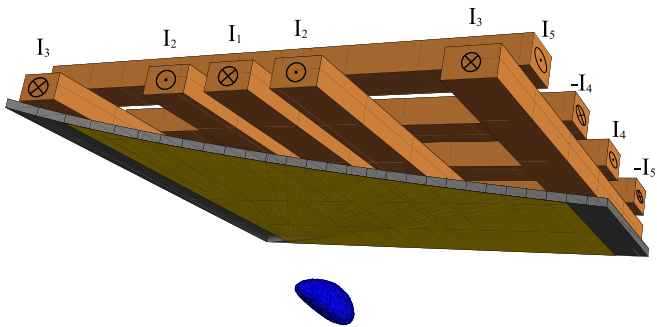}
		\caption{Wire structure to form a hexapole compensated quadrupole field for the operation of a large volume magneto-optical trap. The copper bars indicate the current carrying wires. The sheet indicates the gold coated steel foil that forms the vacuum chamber and allows the chip to be placed outside the vacuum. A surface of equimagnetic field strength is shown in blue, indicating the position of the magneto-optical trap.}
	\label{fig:3d-quad}
\end{figure}

In addition to this two-dimensional quadrupolar field, at least two additional wires in the chip plane are required to produce the three-dimensional quadrupole field required for the MOT. These wires are orthogonal to the five previous ones and carry opposing currents $I_4$ and $-I_4$. Such a configuration suffers from the fact that the two wires do not only provide a quadrupole component in the third, axial direction but also add an offset field pointing in the vertical direction. This offset field shifts the center of the quadrupole trap parallel to the atom chip and has to be avoided. This offset can be compensated by a second pair of wires at a larger distance with opposite currents $-I_5$ and $I_5$. The full three-dimensional field configuration is depicted in Fig.~\ref{fig:3d-quad}.
In principle, the spatial size of the configuration can be chosen at will. The total size can be scaled at the cost of linearly increasing currents to sustain the magnetic field gradients. Within our experiments, we chose a distance of $8$~mm between the atom chip and the center of the trap, generating a large trapping volume wiith a diameter of approx. $8$~mm at reasonable currents. With currents $I_1,I_2,I_3,I_4,I_5=40,-48.5,66.1,48,-29$~A, a three-dimensional, hexapole compensated quadrupole field can be generated with gradients of $11$~G/cm and $6$~G/cm in the two radial directions and $5$~G/cm in the axial direction.

\subsection{Chip design}
\label{chipdesign}

The configuration outlined above was implemented using wires with a rectangular $2.5\times1.5$~mm$^2$ cross section which are able to support the currents with appropriate cooling. These wires were embedded in a copper block according to the optimized wire positions. The outer two of the five longitudinal wires are separated $12.2$~mm from the central wire and the inner ones $3.7$~mm. The four perpendicular wires are $1.7$~mm higher (as shown in Fig.~\ref{fig:3d-quad}) and have a separation of $6.4$~mm. To allow for good thermal contact, the wires were glued into the block with thermally conductive epoxy resin (Fischer Elektronik WLK). The low viscosity of the glue before hardening ensures precise positioning of the wires. To ensure sufficient cooling, the copper block is internally water-cooled close to the bottom facet. 

The wires were bent around the corners of the copper block, thus leading perpendicularly away from the chip plane. These out-of-plane parts of the wires also contribute to the magnetic field of the trap. However due to their larger distance from the center of the MOT, their contribution is well approximated by an offset field. Furthermore, the non-vanishing cross section of the wires leads to a slight change of the field configuration that was optimized by assuming infinitesimally small wires. We performed a full numerical Biot-Savart simulation including these contributions, showing an additional offset field at the intended position of the trap center. The field configuration with the described advantages was recovered by adjusting two currents to $I_3=120$~A and $I_4=53$~A to cancel the offset contributions. The hexapole compensation was almost unaffected. In total, the described setup fulfills the requirements for a chip MOT with a large trapping volume. 

To provide the required currents a single high current power supply (TDK Lamda GEN-7.5-1000) is used. All atom chip wires are supplied in parallel and the current in each wire is controlled by a specifically designed H-bridge controller in a servo-loop, allowing for both current directions. Thus versatile magnetic field configurations with individual currents up to $150$~A and fast modulations up to $6$~kHz can be created.

\section{Experimental setup}
\label{setup}

\begin{figure}[ht]
	\centering
		\includegraphics[width=\columnwidth]{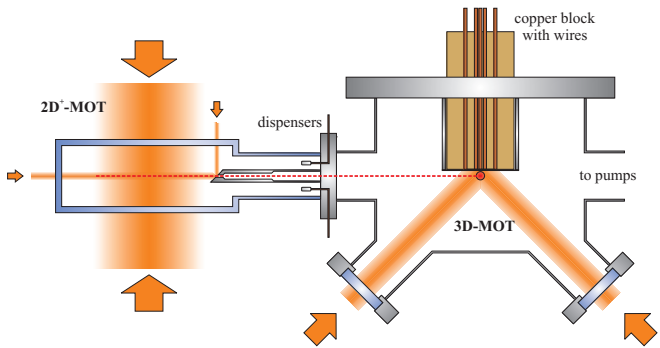}
		\caption{Simplified sketch of the experimental setup, showing the  2D$^+$-MOT and the 3D-MOT. The path of the atomic beam is indicated with a dashed line while the position of the 3D-MOT is marked with a red dot. The reentrant part of the vacuum system that allows for the atom chip to be placed outside the vacuum system and the gold coated steel foil are sketched above the MOT position. The third dimension of the system, which will allow for further transport of the atomic clouds to a shielded region, has been omitted for clarity.}
	\label{fig:setup}
\end{figure}

The experimental setup is designed to generate the mesoscopic chip MOT but it will also allow for a further magnetic transport of the trapped ensemble to an optically shielded region of better vacuum. Thus it will be possible to reload the MOT as soon as the previous ensemble has been transported to the shielded region. To achieve a high cycle rate for this transport a powerful loading scheme for the three-dimensional MOT (3D-MOT) is necessary. This is realized by an additional two-dimensional MOT (2D$^+$-MOT). A sketch of the total setup is shown in Fig.~\ref{fig:setup}. 

\subsection{System layout}

The vacuum system consists of a custom made steel chamber supporting the atom chip for the 3D-MOT and a quartz cell ($6\times6\times15$~cm) for the 2D$^+$-MOT. The quartz cell is separated from the main chamber via a differential pumping stage, formed by a $9.5$~cm long stainless steel tube with an inner diameter that increases from $2$~mm to $8$~mm.

The main chamber is divided into the two regions by a stainless steel plate. The MOT region is maintained at a pressure of  $1\EE{-9}$~mbar  by a  $55$~l/s ion getter pump. The shielded vacuum region is separated from this area by a further differential pumping stage in the stainless steel plate. This will allow for the magnetic transport of the atomic clouds to the shielded region in further experiments. The shielded region is pumped to a pressure below $1\EE{-11}$~mbar by a $150$~l/s ion getter pump and a liquid nitrogen cooled titanium sublimation pump. 

To be able to place the atom chip outside the vacuum system a special reentrant flange (schematically shown in Fig.~\ref{fig:setup}) was designed. While the side walls of this reentrant flange are made of regular steel plates, its bottom is formed by a $500~\mu$m thick nonmagnetic steel foil. It is coated with a $5~\mu$m thick gold layer on the vacuum side and thus also acts as a mirror for the MOT laser beams as indicated in Fig.~\ref{fig:setup}. A finite element simulation showed that the stainless steel foil is not distorted by more than $150~\mu{}$m due to the atmospheric pressure and thus provides a sufficiently good mirror surface. Since the MOT is formed at a distance of about $7$~mm from the gold surface (see Fig.~\ref{fig:2d-quad}) this does not provide any experimental limitations. Indeed this design allows to supply the atom chip with currents and cooling water without the need for vacuum feedthroughs. Hence the contamination of the vacuum system is minimized and very good vacuum conditions can be reached.

\subsection{MOT operation}

The laser light for cooling and trapping of the atoms is derived from two external cavity diode lasers~\cite{Baillard2006} stabilized to the $^{87}$Rb absorption spectrum. To obtain sufficient laser power for the operation of both MOTs, three tapered amplifiers with an output power of $1$~W each are used. Since the laser system and the experimental apparatus are located on separate optical tables, the light is transferred to the experiment with polarization maintaining single mode fibers. 

The 2D$^+$-MOT is operated by two $\sigma$-polarized elliptical light beams with beam diameters of $87$~mm in horizontal and $30$~mm in vertical direction. After passing through the quartz cell, these beams are retro-reflected by $\lambda/4$ plates with a high-reflection coating on their back side. Anti-reflection coatings on the quartz cell's surfaces minimize the power imbalance of the counter-propagating beams. A linearly polarized pushing beam with $5$~mm diameter is used to accelerate the atoms into the main chamber. To reduce the longitudinal velocity of the beam, a toroidal perpendicularly polarized counter-propagating retarding beam is used~\cite{Chaudhuri2006}. This beam is reflected from the polished front surface of the differential pumping tube which is angled $45^\circ$ and thus acts as a mirror for the retarding beam (see Fig.~\ref{fig:setup}).
All beams are detuned by $3.2\,\Gamma$ from the rubidium $^5S_{1/2},F=2\rightarrow{}^5P_{3/2},F=3$ transition and contain repumping light resonant with the $^5S_{1/2},F=1\rightarrow{}^5P_{3/2},F=2$ transition. Each cooling beam has a power of $150$~mW, the pushing and retarding beam have $6.5$~mW and $2.4$~mW respectively. The two-dimensional magnetic quadrupole field for the 2D$^+$-MOT is generated by four coils with $81$ windings each and provides a magnetic field  gradient of $13$~G/cm at a current of $4$~A. By tuning the currents of the coils separately, we adjust the trap center such that the atomic beam passes through the differential pumping stage with minimal losses. The atomic beam is directed slightly below the center of the 3D-MOT's magnetic quadrupole field.

The 3D-MOT is realized in a typical mirror MOT configuration~\cite{Reichel2002}. Two beams with a diameter of $22$~mm and a power of $125$~mW each enter the vacuum system from below at an angle of $45^\circ$. They are reflected from the gold coated foil to provide the radial confinement. The axial confinement is provided by a horizontal laser beam with a diameter of $15$~mm and a power of $71$~mW which  is retroreflected by a mirror inside the vacuum. Similar to the 2D$^+$-MOT, this mirror is a $\lambda/4$ retardation plate with high-reflection coating on the back side. All of the beams are red detuned from the $^5S_{1/2},F=2\rightarrow{}^5P_{3/2},F=3$ transition, and the vertical also contain repumping light.

The trapped atom cloud is investigated using two methods. Firstly, the fluorescence light of the MOT is imaged onto a photo diode for a continuous detection of the total number of atoms. Secondly, two triggered CCD cameras can be used to take fluorescence images along two orthogonal axes to infer density and temperature of an atomic cloud. Both detection methods have been checked for consistency. In total, the presented setup meets the requirements discussed in the introduction.

\section{Mesoscopic chip MOT}
\label{results}

The aim of producing a high-flux source of laser cooled atoms for the preparation of quantum degenerate samples at high cycle rates places stringent requirements on the experimental apparatus. In particular a fast loading time of the 3D-MOT is required while loss mechanisms should be minimal to enable the subsequent transport of atomic clouds. The achieved performance is presented and evaluated with respect to these requirements. In addition, the improved performance due to the hexapole compensation is investigated and the atom number in the 3D-MOT is characterized as a function of the distance from the gold coated foil.

\subsection{MOT loading and decay rate}

Since the 2D$^+$-MOT serves as an atom source for the 3D-MOT, it can be used to analyze the 3D-MOT performance. In the experiments, a magnetic offset field is initially used to shift the center of the 2D$^+$-MOT and thus turn the atomic beam off. To analyze the loading rate of the 3D-MOT, the offset field is quickly removed to start the loading process. The atom number is measured after a certain loading time by fluorescence imaging. To obtain a fluorescence image, all currents and laser beams are switched off $6$~ms before a $100~\mu$s long resonant laser pulse with the vertical MOT beams is applied. For each time, $10$ images were taken to record the fluctuations.

\begin{figure}[ht]
  \centering
    \includegraphics[width=1.0\columnwidth]{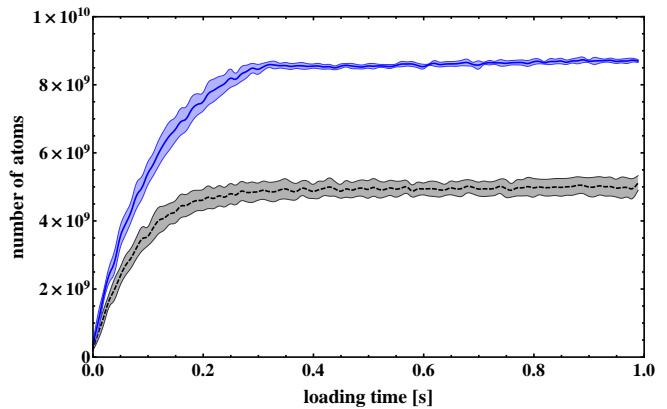}

  \caption{Atom number in the magneto-optical trap as a function of the loading time for a hexapole compensated magnetic field (solid line) and for a similar magnetic field configuration without hexapole compensation (dashed line). The shaded areas corespond to the standard deviation of $10$ evaluated images.}
  \label{fig:loading}
\end{figure}

Figure~\ref{fig:loading} shows the collected number of atoms in the 3D-MOT as a function of the loading time. The data was fitted with an exponential saturation function~\cite{Monroe1990} 

\begin{equation} 
N(t) = N_f(1 - e^{-\frac{t}{T}}) 
\end{equation}

which yields an initial loading rate of $N_f/T=8.4\EE{10}$~atoms/s and a final number of $N_f=8.7\EE{9}$~atoms. Note that the total atom number saturates abruptly at $\approx8\EE{9}$~atoms, leading to an edge in the loading curve. This edge marks the density limit of the MOT~\cite{Park1999} due to reabsorption~\cite{Walker1990} and light-induced collisional losses~\cite{Prentiss1988}. This result can be compared to a field configuration without hexapole compensation, where a quadrupole field with approximately the same gradients is generated by just three wires (the center wire and the two outermost wires). Both the loading rate of $5.9\EE{10}$~atoms/s and the final number of $5.0\EE{9}$~atoms are significantly lower, showing that the hexapole compensation indeed increases the 3D-MOT volume. Thus the hexapole-compensated MOT can be loaded fully within $300$~ms enabling an experimental cycle rate well above 1~Hz, as intended for a high-flux source of laser cooled atoms.

\begin{figure}[ht]
  \centering

   \includegraphics[width=1.0\columnwidth]{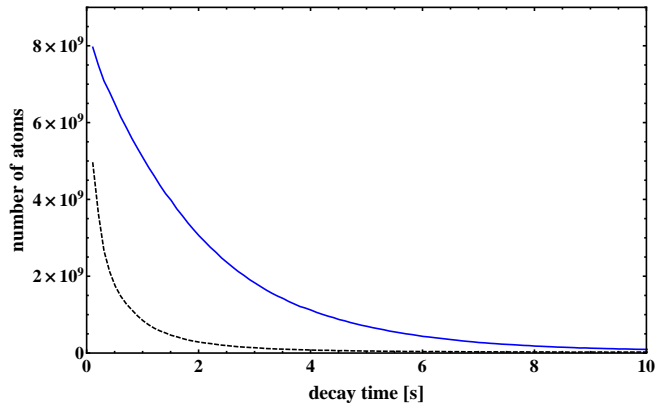}
  \caption{Atom number in the magneto-optical trap as a function of the decay time for a fixed loading time of $500$~ms. Both the case with hexapole compensation (solid line) and without hexapole compensation (dashed line) are shown.}
  \label{fig:decay}
\end{figure}

To measure the decay rate of the 3D-MOT, it was first loaded for $500~$ms such that the atom number is saturated. Then the offset field is applied to the 2D$^+$-MOT to turn off the atomic beam. Figure~\ref{fig:decay} shows the resulting number of atoms as a function of the subsequent hold time. In the configuration with hexapole compensation, the decay is well approximated by an exponential decay function with a lifetime of $2.0$~s, in agreement with the sudden inset of density dependent losses of Fig.~\ref{fig:loading}. This lifetime is fully compatible with losses induced by the background pressure in the MOT region. The situation is strikingly different without hexapole compensation. Due to the distortions of the magnetic field in this configuration the observed lifetime is strongly reduced to $500$~ms and differs from an exponential decay due to additional density dependent losses. This also explains the reduced final number of atoms in Fig.~\ref{fig:loading} since the final number of atoms is determined by the equilibrium between loading and losses.

The total number of atoms exceeds previous results with chip traps~\cite{Reichel1999}, even with basic hexapole compensation~\cite{Wildermuth2004,Hyodo2007} and is even competitive to the $4\EE{10}$~atoms achieved in conventional setups~\cite{Streed2006}.

\subsection{MOT optimization}

To obtain optimal parameters for the operation of the magneto-optical trap we investigated the performance of the system at different configurations for the magnetic field and for the light beams. Within the experimental constraints due to the maximum allowable currents, three representative sets of magnetic field gradients were chosen for this investigation. They correspond to gradients of $5:8:2$~G/cm, $8:12:4$~G/cm and $11:15:5$~G/cm along the three axes of the laser beams. The center of the quadrupole field is at a distance of $7.2$~mm from the gold coated foil in these cases.

\begin{figure}[ht]
  \centering
 
   \includegraphics[width=1.0\columnwidth]{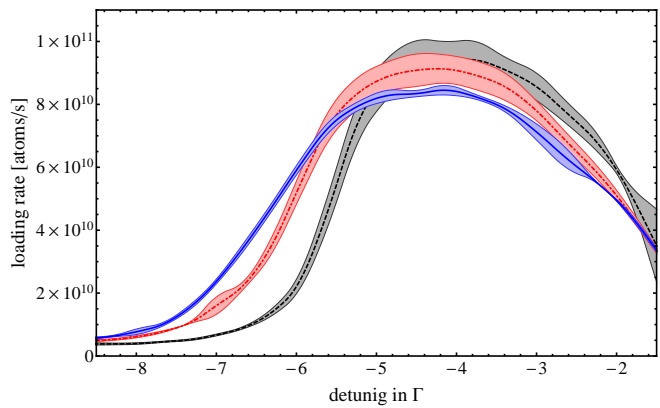}
  \caption{Initial loading rate of the magneto-optical trap as a function of the detuning of the cooling light. The loading rate was determined from a measurement of the number of atoms after $29$~ms loading time. The magnetic field gradients were $5:8:2$~G/cm (black line), $8:12:4$~G/cm (red line) and $11:15:5$~G/cm (blue line) along the three axes of the light beams. The shaded areas corespond to the standard deviation of $10$ evaluated images.}
  \label{fig:rate-detuning}
\end{figure}

Figure~\ref{fig:rate-detuning} shows the loading rate of the 3D-MOT as a function of the cooling laser detuning for these magnetic field gradients. The loading rate is only weakly affected by the chosen magnetic field gradient and rates above $8\EE{10}$~atoms/s are obtained for all configurations at detunings around $4.5~\Gamma$. This insensitivity of the loading rate reflects the finite flux of trappable atoms from the 2D$^+$-MOT.

\begin{figure}[ht]
  \centering
    \includegraphics[width=1.0\columnwidth]{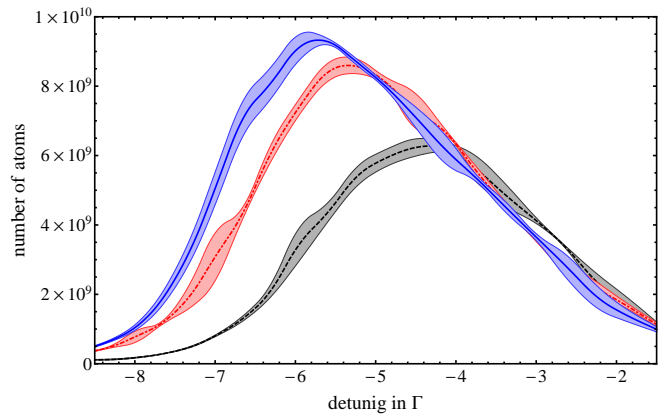}
  \caption{Final number of atoms in the magneto-optical trap after $500$~ms loading time as a function of the detuning of the cooling light. The magnetic field gradients were $5:8:2$~G/cm (black line), $8:12:4$~G/cm (red line) and $11:15:5$~G/cm (blue line) along the three axes of the laser beams. The shaded areas corespond to the standard deviation of $10$ evaluated images.}
  \label{fig:number-detuning}
\end{figure}

In Fig.~\ref{fig:number-detuning} the total number of atoms is shown as a function of the cooling laser detuning for the indicated magnetic field gradients. The performance clearly improves as the magnetic field gradient is increased and the maximum shifts towards larger detunings~\footnote{The use of magnetic field gradients above $11:15:5$~G/cm did not show significant improvements.}. Interestingly, these maxima are reached despite decreased loading rates. This effect can be attributed to the decreased reabsorption rate~\cite{Walker1990} of scattered photons at larger detunings.

These results lead to the optimal parameters of $11$~G/cm, $15$~G/cm and $5$~G/cm for the magnetic field gradients at the detuning of $-5.2~\Gamma$ for the cooling light which provides large atom numbers as well as high loading rates. It also demonstrates that the described configuration is robust against slight fluctuations of the parameters.

\begin{figure}[ht]
	\centering
		\includegraphics[width=\columnwidth]{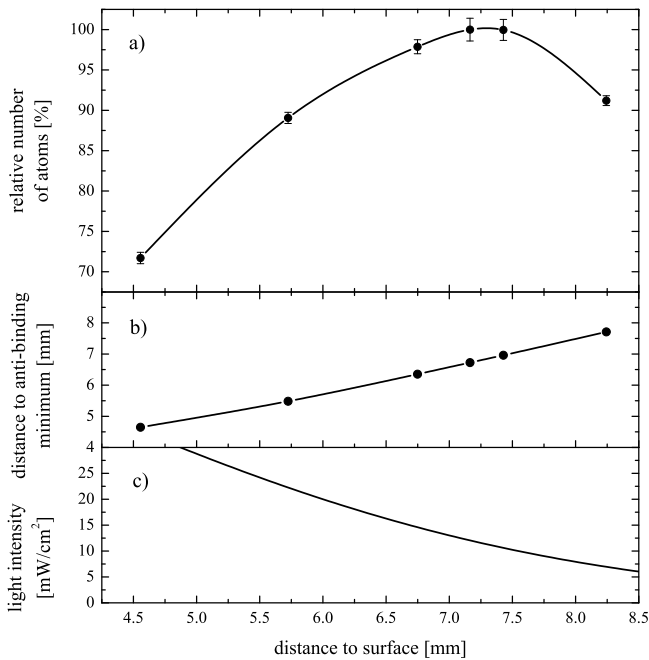}
	\caption{(a) Relative number of atoms as a function of the distance from the atom chip. The gradients were kept at $11\pm 0.6$, $15\pm 0.02$ and $5\pm 0.06$~G/cm within the measurements. The error bars represent the standard deviation of $50$ evaluated images. (b) The distance between the anti-binding minimum and the trap center shows the decreasing spatial size of the MOT close to the surface. (c) At larger distances from the surface, the light intensity of the vertical MOT beams decreases and thus creates an optimal trap center.}
	\label{fig:distance}
\end{figure}

To investigate the influence of the position of the MOT, the number of atoms was recorded as a function of the distance to the gold coated steel foil as shown in Fig.~\ref{fig:distance} (a). The center of the quadrupole field was shifted by adjusting the currents of the outermost wires while preserving the gradients of the trap. The 3D-MOT shows an optimal performance at a distance of $7.2$~mm from the surface corresponding to a distance of $8.6$~mm from the center of the longitudinal wires. Above and below this position, the benefits of the hexapole compensation are reduced since the wire structure was optimized for this distance. In addition, the maximum size of the atomic cloud is limited by the anti-binding minimum shown in Fig.~\ref{fig:2d-quad}. Figure ~\ref{fig:distance} (b) indicates that the anti-binding minimum approaches the trap center for smaller distances to the surface. On the other hand, the lower atom number for larger distances can be explained by the reduction of laser intensity resulting from the mirror MOT configuration, as shown in Fig.~\ref{fig:distance} (c). The optimal measured MOT position thus corresponds to the calculated position, validating the presented optimization procedure for the atom chip.

In order to check the applicability of the MOT as a high flux source of atoms, the minimal temperature after an optical molasses phase was also investigated. A molasses of $3.8$~ms was applied by detuning the cooling beams by $-18.7~\Gamma$ in the absence of magnetic fields. A temperature of $25~\mu$K was recorded after ballistic expansion of $25$~ms, providing a good starting point for further magnetic trapping.

\section{Conclusion and Outlook}
\label{outlook}
In summary, we have presented a mesoscopic atom chip design for a magneto-optical trap. A millimeter sized wire structure provides a hexapole-compensated quadrupole field with a large trapping volume. By using a beam of pre-cooled atoms, fast loading rates of $8.4\EE{10}$~atoms/s and large ensembles with $8.7\EE{9}$~atoms were obtained. The mesoscopic structures are situated outside the vacuum and can thus easily be replaced and modified for different scientific objectives.

The described setup will be used to build a high flux source of quantum degenerate atoms. Since no external fields are required to generate the trapping potential, the structures can be used to simultaneously trap and transport multiple atom clouds at different positions of the chip. In particular the atoms can be trapped in a mesoscopic magnetic conveyor belt~\cite{2Haensel2001} and quickly transferred into a shielded vacuum region. Here, at least five clouds will be cooled simultaneously by spatial evaporation~\cite{Haber2003,Roos2003}. The strong confinement in the chip traps enables evaporation times below $5$~s, yielding a cycle rate of $1$~Hz. Such a rapid source of large Bose-Einstein condensates will pave the way towards competitive atom interferometry with quantum degenerate ensembles.

We acknowledge support from the Centre for Quantum Engineering and Space-Time Research QUEST and the Danish National Research Foundation Center for Quantum Optics (QUANTOP).

\bibliography{Joellenbeck1}

\end{document}